# Prompt photon, Drell-Yan and Bethe-Heitler processes in hard photoproduction

P. J. Bussey[a], B. Levtchenko[b] and A. Shumilin[b]

[a] Department of Physics and Astronomy, University of Glasgow, Glasgow G12 8QQ, UK
[b] Institute of Nuclear Physics, Moscow State University, 119899 Moscow, Russia

**Abstract:** We present prospects and requirements for the study of hard photon processes which generate high $p_T$ photons in the final state, and processes which generate Drell-Yan lepton pairs.

## 1 Introduction

The ZEUS and H1 experiments at HERA have published a variety of studies on hard processes in quasi-real photoproduction. In lowest order (LO) QCD, an incoming photon (or parton from the photon) interacts with a parton from the proton, and two outgoing high $p_T$ partons emerge in the final state, which can then hadronise to give rise to two observable jets. Two major classes of LO process are defined: direct, in which the entire photon interacts in the QCD subprocess, and resolved, in which the photon is a source of partons one of which interacts. A major objective is to determine the parton densities in the photon and the proton, the latter complementing the many measurements which have been made in DIS processes.

A problem concerning measurements of this type concerns the effects of final state QCD radiation, or higher order QCD effects in general. These are reduced if we can measure processes in which the emerging particles from the QCD subprocess are not themselves subject to QCD effects. Two classes of process which have been used in this way in hadron collisions are those in which high-$p_T$ photons are produced, i.e. so-called "prompt" photons, and those in which quark-antiquark annihilation gives rise a pair of leptons, known as Drell-Yan processes. Fig. 1 illustrates the different types of process. In "dijet" processes, both final state particles in the basic diagram are quarks or gluons, while the others involve a photon or a pair of leptons.

In $p\bar{p}$ collisions at Fermilab, prompt photon processes provide a way to study the gluon content of the proton [1, 2]. At HERA, the accessible kinematic range restricts the main sensitivity to the quark content of the photon [3–6], together with the quark and gluon contents of the proton. The particular virtue of prompt photon processes is that the observed final-state photon emerges from the QCD process directly, without the intermediate hadronisation that accompanies the observation of a quark or gluon through the means of a final state jet. This, together with the availability of NLO calculations [7, 5], makes such processes attractive in providing the prospect of a relatively clean technique for studying QCD. On the other hand, the cross sections are substantially lower than those of dijet processes.

|           | DIJET | PROMPT PHOTON | BETHE-HEITLER DRELL-YAN |

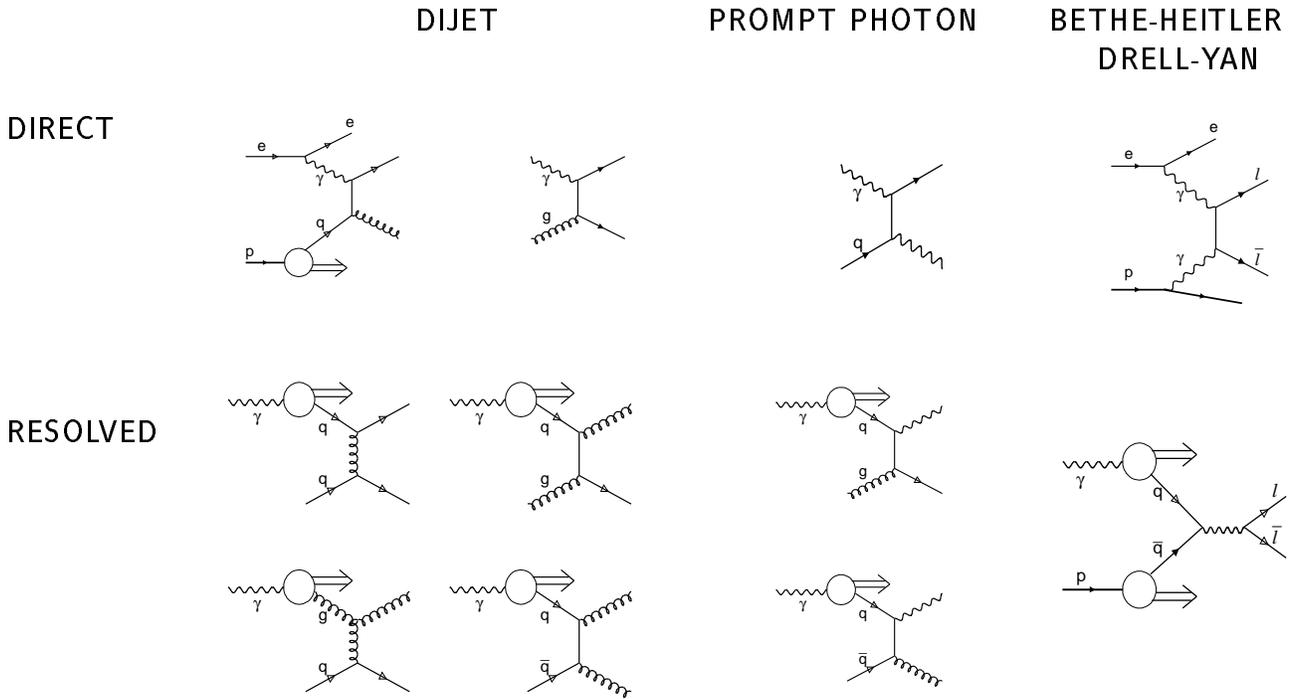

Figure 1: *Main LO diagrams for direct dijet and prompt photon processes in hard photoproduction, and examples for resolved processes. Diagrams for Bethe-Heitler and LO Drell-Yan processes are also shown. Broad arrows represent photon or photon remnants.*

Drell-Yan (DY) processes in photoproduction are by definition resolved at LO. There is no exactly parallel class of direct events, but a background comes from Bethe-Heitler (BH, "photon-photon") processes, as illustrated in Fig. 1. (A further BH diagram, not shown, has an inelastic excitation of the proton at the lower vertex.) Both BH photons interact in a direct way in producing the lepton pair. Direct-resolved and resolved-resolved photon photon interactions can also occur and give hadronic final states. A higher order "direct" DY diagram can also be drawn in which the photon remnant is replaced by a high-$p_T$ $\bar{q}$ or $q$ [8].

The study of lepton pair production through DY processes is of interest because it can provide a further alternative way to measure the parton densities of the photon and the proton, testing perturbative QCD and the determination of the running coupling constant. It is also an important background process for other production mechanisms of lepton pairs, such as $J/\Psi$ and $\Upsilon$ decays. The twist-two chirality violating proton structure function $h_1(x)$ can be measured in the DY reaction when both beam and target are transversely polarized. In the framework of the nuclear program at HERA the DY reaction will allow investigation of the violation of charge symmetry in the valence quark distributions of the nucleon at large $x$, and tests of $SU(2)$-flavour symmetry breaking.

In 1971 Jaffe [9] suggested that the photon structure function could be determined through the DY process in photoproduction experiments. Since then, several other authors have studied lepton pair production theoretically using the QCD formalism at kinematic conditions at fixed target experiments [10], as well as $ep$ colliders such as HERA [11]. Much attention was paid to find proper kinematic variables and to determine a kinematic region where the major contribution from the DY leptons would not be dominated by the background processes. All of them, however, reached the not very comforting conclusion that, because of the overwhelming Bethe-Heitler background, photoproduction of lepton pairs from a proton target is not an adequate way of measuring the photon structure function through the DY effect.

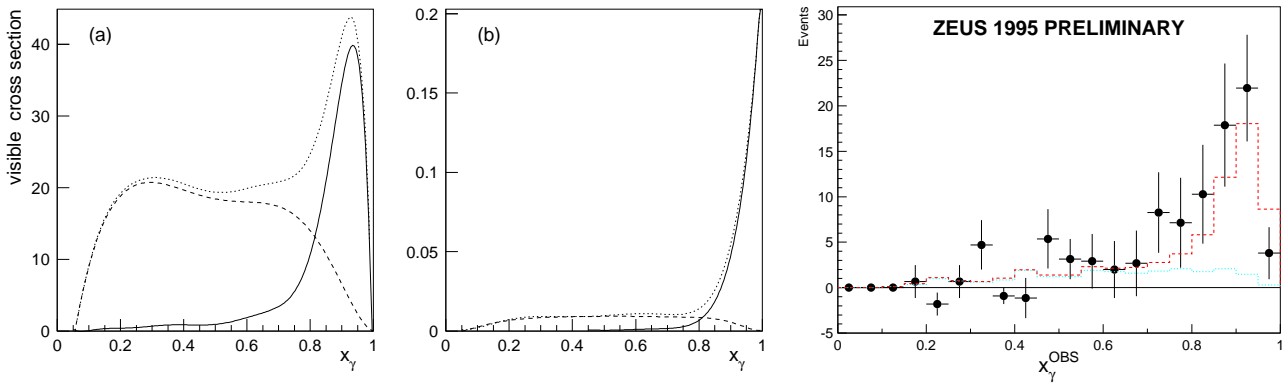

Figure 2: *Distributions of $x_\gamma$ [6] for (a) dijet events, (b) prompt photon events as observed in an idealised HERA detector. Solid curve = direct contribution; dashed curve = resolved; dots = total. (c) Preliminary ZEUS results [15]: dashes = resolved + direct, dots = resolved.*

The planned HERA upgrade, however, will increase the integrated luminosity by two orders of magnitude and allow stricter selection criteria to separate the processes of interest. The conclusion of [11] for HERA was based on a comparison of inclusive spectra ($p_T$ and rapidity), calculated for $4\pi$ geometry without taking into account the acceptances of the H1 or ZEUS detectors. One should therefore make an effort to analyse the Drell-Yan and background processes using a more precise approach to the experimental conditions. Results presented below show that it is possible to find kinematic criteria under which the Bethe-Heitler background is totally suppressed.

## 2  Prompt photon processes

The quantity $x_\gamma$ is defined as the fraction of the photon energy taking part in the hard QCD subprocess, and is a powerful tool in characterising high energy photoproduction processes [13, 14]. For direct processes its value is by definition unity. Experimentally, its distribution is expected to differ markedly between dijet processes and prompt photon processes. An "observed" value of $x_\gamma$ may be evaluated, using the definitions of [13] and [6], as $\sum_{jets}(E - p_Z)/\sum_{event}(E - p_Z)$, summing over particles (or calorimeter cells) in the jets (or the jet plus the photon) and in the whole event, for dijet events or prompt photon events as appropriate.

Figure 2 shows the difference between the distributions of the "observed" value of $x_\gamma$ between dijet events and events with a prompt photon and an accompanying high-$E_T$ jet, subject to the typical kinematic constraints of a HERA experiment. High $E_T$ jets have been reconstructed ($E_T > 5$ GeV) from the four-vectors of HERWIG simulations of direct and resolved events, with jets and prompt photons (also $E_T > 5$ GeV) accepted in the pseudorapidity range $-1.5 \leq \eta \leq 1.7$. First experimental results from ZEUS have been reported at the Rome DIS 96 Workshop [15]. In round figures, an integrated luminosity of 6 pb$^{-1}$ gives approximately 50 direct events with a prompt photon in the ZEUS barrel calorimeter and 30 resolved events, an accompanying jet also being observed.

The Direct Compton diagram gives a good measurement of the quark content of the proton, but an even more important aim is to measure the photon structure by means of the resolved events. With an integrated luminosity of 1000 pb$^{-1}$ we may thus expect to record around 5000 resolved events and 7500 direct events with a prompt photon and a jet. This should suffice to give a reasonable measurement of the photon quark density and distinguish between present models which currently give cross sections differing among themselves by typically 10-20% [6]. Such measurements would be noticeably degraded if the total integrated luminosity were less, say, by a factor of 4. Use of photons detected in the rear calorimeter would give a small improvement, but not in the numbers of resolved events. To improve the resolved statistics we would require a better understanding of photon detection in the forward calorimeter. This is technically difficult, however, and its viability needs further study.

An alternative possibility would be to use inclusive prompt photon distributions. These [4] appear to give a better sensitivity to the photon structure, different models varying by as much as 40%. The inclusive cross section for prompt photons within a pseudorapidity range of $|\eta| \leq 1$ is given by [4] as 24-34 pb for the resolved contribution, with a similar figure for the direct. An integrated luminosity of 1000 pb$^{-1}$ will then give around 30k events in each category, the gain coming by an avoidance of the need to detect the jet. However to take advantage of these statistics, it would still be necessary to distinguish somehow between direct and resolved events. One way to do this is to model the photon remnant in terms of energy detected (outside any high $E_T$ jets) in the rear regions of the detector, with the aim of reconstructing $x_\gamma$ approximately for each event. First investigations have been made [16] but the technique is clearly more difficult than if the outgoing jet is detected. For this reason, it seems necessary at present to aim for an $x_\gamma$ measurement, for which the highest possible luminosity is required.

## 3 Drell-Yan and Bethe-Heitler processes

In our present analysis, $e^+e^-$ and $\mu^+\mu^-$ production in $ep$ collisions ($E_e = 26.7$ GeV, $E_p = 820$ GeV) was simulated by PYTHIA5.7 (for DY leptons) and LPAIR [17] (for BH leptons). This allows us easily to apply all the necessary cuts to account for detector resolution and detector geometry. The results presented here are correspond to the geometry of the ZEUS detector and are based on samples of $2 \cdot 10^4$ events (for DY pairs) and $10^5$ events (BH pairs).

To study the effects of different selection criteria on the background suppression, cuts on the the polar angle $\theta_l$ and energy $E_l$ of produced leptons were defined as follows: $1L_e \equiv 2.2° < \theta_e < 176.5°$; $1L_\mu \equiv 5° < \theta_\mu < 170°$; $2L_e \equiv 2.2° < \theta_{e^+}, \theta_{e^-} < 176.5°$; $2L_\mu \equiv 5° < \theta_{\mu^+}, \theta_{\mu^-} < 170°$; $1E \equiv E_l > 1$ GeV, $1E5 \equiv E_l > 5$ GeV, $2E \equiv E_{1,2} > 1$ GeV. These conditions correspond to lepton acceptances in the ZEUS detector. Two types of trigger were considered: a) "tagged", requiring the detection of the scattered electron ($5 < E_{e'} < 25$ GeV) at very small angles, thereby limiting $Q^2$ to less than 0.02 GeV$^2$, and b) "untagged", requiring the absence of a detected scattered electron in the main rear calorimeter ($Q^2 < 4$ GeV$^2$).

The total "untagged" and "tagged" cross sections for photoproduction of DY pairs with masses $M_{l^+l^-} > 1$ GeV were found to be 84 pb and 25 pb respectively.

Table 1 presents numbers of events with electron pairs passing the different combinations of cuts defined above. From the last column of the table it can be seen that the requirement of detecting at least one of the BH electrons in the calorimeter ($1L_e \otimes 1E$) dramatically reduces the number of BH events passing this cut. With simultaneous detection of both electrons ($2L_e \otimes 2E$) only about 12% of the DY electron pairs survive, but the BH background is totally eliminated. The ZEUS muon detector system has a smaller angular acceptance and so the number of DY muon pairs passing the $2L_\mu \otimes 2E$ cuts is 1548 events of 20 000.

| TRIGGERS | PYTHIA $Q^2 < 4\,GeV^2$ | PYTHIA $Q^2 < 0.02\,GeV^2$ | PYTHIA $Q^2 < 0.02\,GeV^2$ $5\,GeV < E_{e'} < 25\,GeV$ | LPAIR |
|---|---|---|---|---|
| No Cuts | 20 000 | 20 000 | 7 125 | $10^5$ |
| 1L | 14 301 | 14 099 | 5 903 | 54 569 |
| 2L | 12 475 | 12 248 | 3 974 | 44 649 |
| 1E | 15 560 | 15 589 | 5 221 | 5 801 |
| 1E5 | 8 874 | 9 058 | 2 209 | 1 050 |
| 1L⊗1E | 9 824 | 9 693 | 3 997 | 8 |
| 2L⊗2E | 5 771 | 5 609 | 2 439 | 0 |

Table 1: *Numbers of events with photoproduction of an electron pair (PYTHIA 5.7 for DY and LPAIR for BH) under different trigger conditions.*

Figure 3 illustrates these results and shows the (pseudo)-rapidity distributions of DY and BH leptons (i.e. electrons + muons) under different trigger conditions. The final plot summarizes the results of our analysis; at an integrated luminosity of 10 pb$^{-1}$, with the 2L⊗2E cuts in the ZEUS detector, about 40 events can be detected with DY electron pairs and 25 events with DY muon pairs. With an integrated luminosity of 1000 pb$^{-1}$ these data samples will increase by two orders of magnitude, which should just be sufficient to enable the photon structure to be investigated.

A further feature which could be used to distinguish DY from BH events is the presence of a photon remnant. This can be quantified once again by evaluating $x_\gamma$ (in terms of the two leptons instead of jets). The BH events will have $x_\gamma \approx 1$; $x_\gamma$ must be evaluated in any case for a study of the photon structure, and a cut to remove high $x_\gamma$ events should help to remove the BH background and perhaps allow other conditions to be loosened. We have not had time to investigate this question further, but it is clear that a number of possibilities exist for studying DY pairs at HERA. All require the highest luminosities that can be obtained.

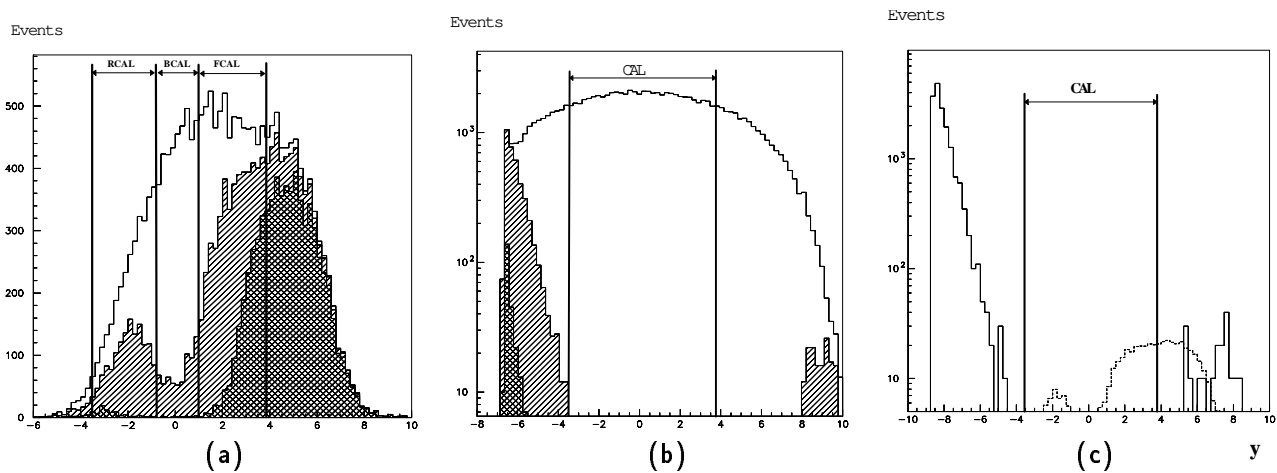

Figure 3: *Pseudo-rapidity distributions of (a) DY leptons for all events (open histogram), and the subsets with applied cuts 2E (shaded, light) and 2E5 (shaded, heavy), (b) BH leptons with the same cuts as above, (c) leptons passing the 2L⊗2E cuts with an integrated luminosity of 10 pb$^{-1}$. The solid and dashed histograms correspond to BH and DY leptons respectively. The acceptance of the ZEUS calorimeter is indicated.*

## 4 Conclusions

To study the photon structure through prompt photon and Drell-Yan processes will require the highest integrated luminosities that HERA can deliver. We support a full upgrade to an integrated luminosity of 1000 pb$^{-1}$.